\documentclass{ws-ijmpe}

\begin{document}

\markboth{M. Warda, A. Staszczak, L. Pr\'ochniak}
{Comparison of Skyrme and Gogny results in light Hg isotopes}

\catchline{}{}{}{}{}

\title{COMPARISON OF SELF-CONSISTENT SKYRME AND GOGNY CALCULATIONS FOR LIGHT Hg ISOTOPES}

\author{\footnotesize M. WARDA$^a$, A. STASZCZAK$^{a,b,c}$, L. PR\'{O}CHNIAK$^a$}

\address{$^a$Department of Theoretical Physics, Institute of
Physics, Maria Curie-Sk{\l}odowska University,\\
pl.\ M.\ Curie-Sk{\l}odowskiej 1, 20-031 Lublin, Poland\\
$^b$Department of Physics, University of Tennessee, Knoxville,
TN 37996, Knoxville, USA\\
$^c$Physics Division, Oak Ridge National Laboratory, P.O.Box
2008, Oak Ridge, TN 37831, USA\\
}

\maketitle

\begin{history}
\received{(received date)}
\revised{(revised date)}
\end{history}

\begin{abstract}
The ground-state properties of neutron-deficient Hg isotopes have been investigated by the constrained self-consistent Hartree-Fock-Bogoliubov approach with the Skyrme and Gogny effective forces. In the case of the Skyrme interaction we have also applied the Hartree-Fock+BCS model with the state-dependent $\delta$-pairing interaction. Potential energy surfaces and pairing properties have been compared for the both types of forces.
\end{abstract}

\section{Introduction}

The self-consistent mean-field Hartree-Fock-Bogoliubov (HFB) method is one of the standard approaches in nuclear structure and low-energy dynamics research.\cite{RS80} In this paper, we compare two self-consistent models: the HFB framework with the Skyrme interaction\cite{Skyrme} and the HFB theory with the Gogny force.\cite{DG} We have tested these two models with various parameter sets. For calculations we have chosen the commonly used SkM$^*$ and SLy4 parameterizations of the Skyrme force.  We have also performed calculations with the SkM$^*$ parameterization the Hartree-Fock+BCS (HF+BCS) model. This has been done in order to check the impact of a non-self-consistent BCS scheme on the results in the case of nuclei far from the stability line. The acknowledged D1S parameter set of the Gogny force has been set together with recently developed D1N and D1M sets.

The neutron-deficient nuclei, in the vicinity of the neutron mid-shell at $N$=104 and near the $Z$=82 shell closure, exhibit a variety of coexisting shapes (see, e.g., Refs.\cite{Wood92,Andr00,Julin01}). The potential energy surfaces (PESs) of nuclei from this region show complex structures around their ground states. This brings about a stringent restriction on the validity of various nuclear models and forces parameters. We have chosen light $^{178,180,182,184}$Hg isotopes for tests of the mean-field models with the Skyrme and Gogny forces.

\section{Calculation details}

Calculations with the Skyrme interaction have been carried out using a symmetry-unrestricted code \textsc{HFODD} (v.2.38j)\cite{hfodd1-3,hfodd4-5,hfodd6},
that solves the self-consistent HF+BCS and HFB equations in the Cartesian deformed harmonic-oscillator basis. In the calculations, for the basis we have taken the lowest 1140 stretched single-particle states originated from the lowest 31 and 26 major oscillator shells for the HF+BCS and HFB approach, respectively.

In the \emph{ph}-channel we have applied the Skyrme energy density functional with the SkM$^*$ Ref.\cite{SkM*} and SLy4 Ref.\cite{SLy4} parameterizations. In the \emph{pp}-channel we have used the MIX variant of state-dependent $\delta$-interaction (see, \textit{e.g.,} Ref.\cite{pairing})
\begin{displaymath}
\begin{array}{rcl}
V^{n/p}_{\delta}\left(\vec{r}_{1},\vec{r}_{2}\right)&=&
V^{n/p}_{0}\left[ 1-\frac{1}{2} \left(\frac{\displaystyle\rho \left({\textstyle
\frac{\vec{r}_{1}+\vec{r}_{2}}{2}} \right)}
{\displaystyle\rho_{0}}\right)^{\alpha}\right]
\delta\left(\vec{r}_{1}-\vec{r}_{2}\right),
\end{array}
\label{1}
\end{displaymath}
where $\rho_{0}=0.16\,\mbox{fm}^{-3}$ and $\alpha=1$.

The pairing strengths for neutrons/protons
$V^{n}_{0}=-372.0$~MeV, $V^{p}_{0}=-438.0$~MeV (SkM$^*$-BCS),
$V^{n}_{0}=-268.9$~MeV, $V^{p}_{0}=-332.5$~MeV (SkM$^*$-HFB),
and $V^{n}_{0}=-308.0$~MeV, $V^{p}_{0}=-343.4$~MeV (SLy4-HFB)
have been adjusted to reproduce the experimental pairing gaps in $^{252}$Fm.
As we deal with contact interactions, for the HF+BCS approach we have used a finite pairing-active space defined by including $N/Z$ lowest single-particle states for neutrons/protons, while for summing up contributions of the HFB quasi-particle states to density matrices the cutoff energy of 60~MeV has been used.

\begin{table}[htb]
\tbl{Gogny parameters in D1S, D1N and D1M parameters sets}
{\begin{tabular}{l|rrr}
\toprule
parameter      &       D1S &        D1N &        D1M \\
\colrule
$W_1$ [MeV] & -1 720.30 &  -2 047.61 & -12 797.57 \\
$B_1$ [MeV] &  1 300.00 &   1 700.00  &  14 048.85 \\
$H_1$ [MeV] & -1 813.53 &  -2 414.93 & -15 144.43 \\
$M_1$ [MeV] &  1 397.60 &   1 519.35 &  11 963.89 \\
$\mu_1$ [fm]&       0.7 &        0.8 &        0.5 \\
\colrule
$W_2$ [MeV] &   103.639 &     293.02 &     490.95 \\
$B_2$ [MeV] &  -163.483 &    -300.78 &    -752.27 \\
$H_2$ [MeV] &   162.812 &     414.59 &     675.12 \\
$M_2$ [MeV] &  -223.934 &    -316.84 &    -693.57 \\
$\mu_2$ [fm]&       1.2 &        1.2 &        1.0 \\
\colrule
$t_0$ [MeV fm$^{3(1+\gamma)}$]
            &   1 390.6 &   1 609.46 &   1 562.22 \\
$x_0$       &         1 &          1 &          1 \\
$\gamma$    &     $1/3$ &      $1/3$ &      $1/3$ \\
$W_{LS}$[MeV fm$^5$]
            &       130 &      115.0 &     115.36 \\
\botrule
\end{tabular}}
\label{parameters}
\end{table}

The most popular parameter set of the Gogny force D1S was fitted almost twenty years ago.\cite{B} Recently, two new parameter sets have been proposed: D1N\cite{C} and D1M.\cite{G} These parameterizations have been developed with special care about accurate reproduction of nuclear matter properties and dependence of symmetry energy on nuclear density. The D1M set has been fitted to reproduce over 2000 experimental masses after including beyond-mean-field quadrupole corrections. All parameters of the considered Gogny forces are given in Table \ref{parameters}.

The Gogny calculations\cite{W} have been performed in the deformed axial basis with $N_z=22$ oscillator shells in $z$ direction and $N_\perp=15$ in perpendicular directions. Oscillator lengths have been optimized in every point of calculations.

\section{Results}

\begin{figure}[h]
\centerline{\psfig{file=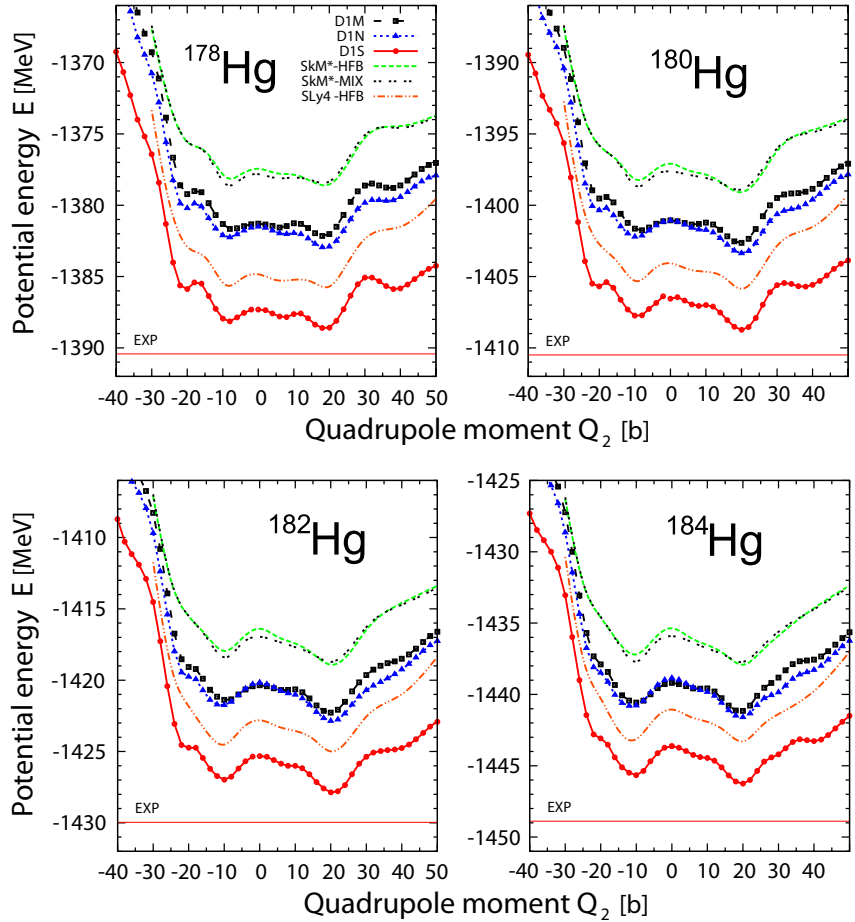,width=0.9\textwidth}}
\caption{(Color online) Potential energy curves from the Skyrme and Gogny self-consistent models restricted to axially symmetric shapes as a function of Q$_2$ mass moment calculated for $^{178,180,182,184}$Hg isotopes (for details, see text).}
\label{fig1}
\end{figure}

In Fig.~\ref{fig1} we present potential energy curves of four neutron-deficient isotopes $^{178,180,182,184}$Hg calculated for axially symmetric shapes for moderate deformations ($|Q_{2}| < 50$~b). The energies have been calculated with the use of two different effective forces: Skyrme and Gogny. In the case of the HFB-Gogny approach we have used the DS1, D1N and D1M parameter sets. The SkM$^*$ and SLy4 Skyrme parameterizations have been applied to the HFB-Skyrme framework (curves marked as SkM$^*$-HFB and SLy4-HFB in Fig.~\ref{fig1}). In addition, the SkM$^*$ set has been applied to the HF+BCS model with the MIX paring $\delta$-interaction (SkM$^*$-MIX). The potential energy profiles shown in Fig.~\ref{fig1} include only the mean-field energies with projection onto good center-of-mass momentum. Other beyond mean-field corrections, e.g., a rotational energy correction, have been neglected.
\begin{figure}[ht]
\centerline{\psfig{file=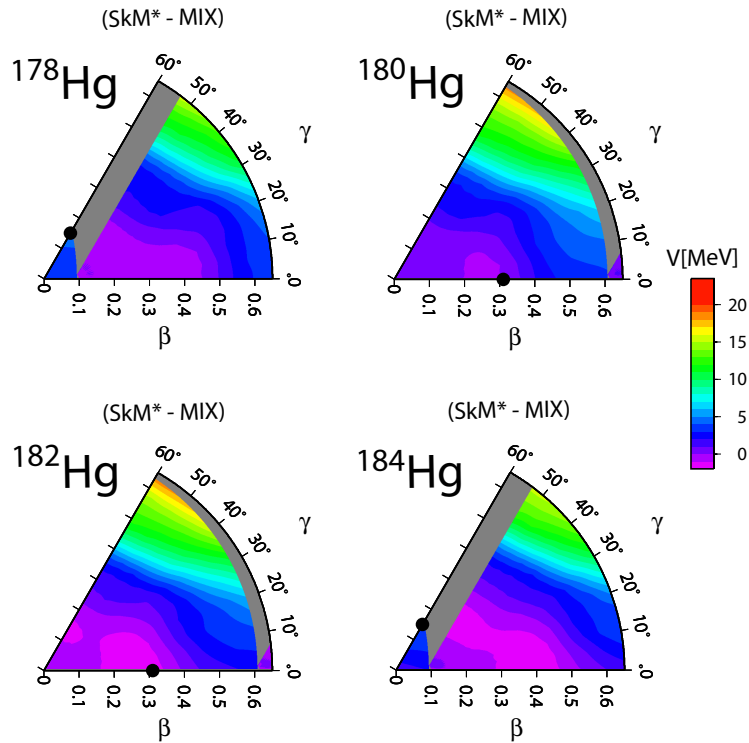,width=0.8\textwidth}}
\caption{(Color online) Relative potential energy surfaces calculated in the SkM$^*$-MIX model for $^{178,180,182,184}$Hg in $\beta-\gamma$ plane.}
\label{fig2}
\end{figure}

In spite of the fact that we have used various mean-field approaches, results obtained with all parameter sets are quite similar. First of all, one can easily find five minima in all presented isotopes and models: two at the oblate side ($Q_2\approx-20$ and $-10$~b) and three with prolate deformation ($Q_2\approx5$, $20$ and $40$~b). The most pronounced prolate minimum is localized at $Q_2\approx20$~b and the HFB models predict ground state there. Nevertheless, in all considered nuclei, the deepest oblate minimum at $Q_2\approx-10$~b is not more than 1.2~MeV higher for the Gogny forces, and even less for the Skyrme forces. It is worth to note that the deformed prolate character of the ground state is best seen in $^{180}$Hg and $^{182}$Hg for all models. This is in accordance with experimental results published recently\cite{grahn09}. In $^{178}$Hg and $^{184}$Hg the HFB models predict oblate minimum on 0.5~MeV above the prolate ground state, whereas the SkM$^*$-MIX model indicate oblate ground state. The other minima ($Q_2\approx-20$, $5$ and $40$~b) are not distinct, especially in isotopes $^{182}$Hg and $^{184}$Hg where they convert into shoulders only.

\begin{figure}[ht]
\centerline{\psfig{file=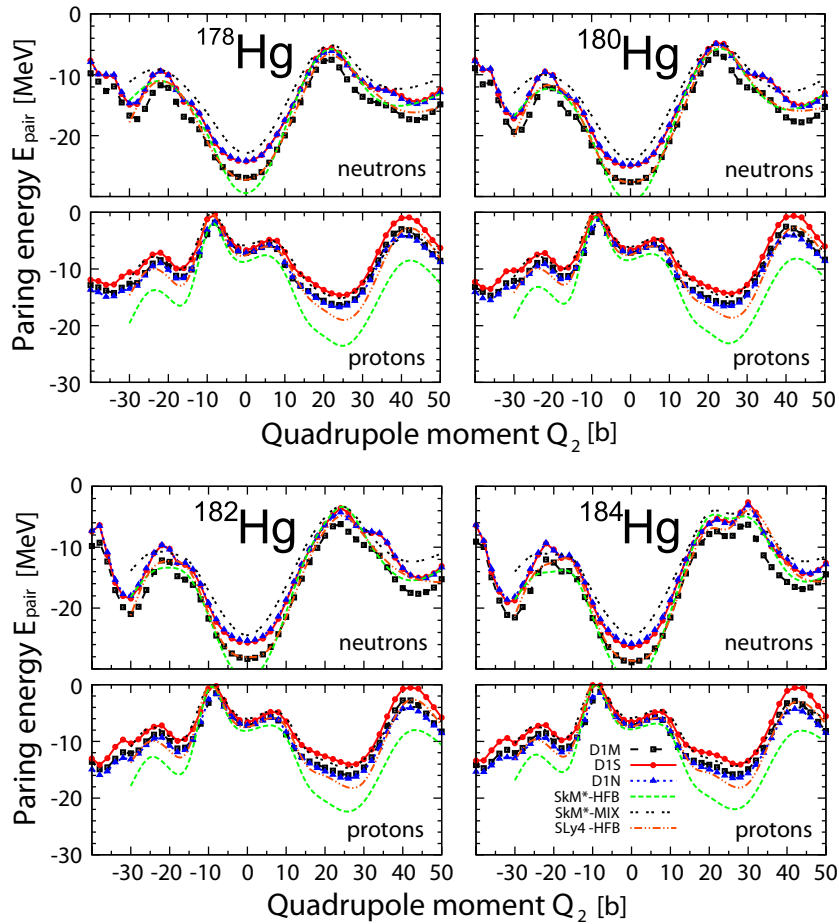,width=0.9\textwidth}}
\caption{(Color online) Neutron and proton pairing energy for isotopes and models presented in Fig. 1.}
\label{fig3}
\end{figure}

The PESs calculated with the D1N and D1M Gogny forces lie higher by about 6~MeV than the values obtained with the D1S force. The binding energies in these two parameterizations coincide for spherical shapes and do not differ more than 1~MeV for small quadrupole deformations. Inclusion of the beyond-mean-field quadrupole corrections in the D1M set improve substantially agreement between theoretical and experimental masses.\cite{G} However, in the case of four neutron-deficient Hg isotopes investigated here, the D1S parameter set gives the binding energies closest to the experimental values, marked in each panel by horizontal lines.

The PESs calculated with the Skyrme SLy4-HFB model are shifted up by around 3~MeV in relation to the values obtained with the D1S Gogny force. In both the SkM$^*$-HFB and SkM$^*$-MIX models the calculated PESs are almost identical, however, their binding energies overestimate the experimental values by around 10~MeV. This is the worst agreement for all models investigated here, which indicates that the SkM$^*$ set fails in prediction of absolute experimental data for nuclei in this region.

For better understanding the relations between minima presented in Fig. 1, we have performed calculations with broken axial symmetry. Fig. 2 shows the relative potential energy surfaces calculated in the SkM$^*$-MIX model for $^{178,180,182,184}$Hg in the $\beta-\gamma$ plane. For $^{180,182}$Hg isotopes, one can find shallow prolate global minima at $\beta\approx0.3$ that correspond to the ground states at $Q_2=20$~b in Fig. 1. However, in the case of $^{178,184}$Hg isotopes, the SkM$^*$-MIX model suggests even more shallow oblate ground state minima at $\beta\approx0.15$ ($Q_2=-10$~b). In the non-axial calculations these oblate minima have appeared to be extremely soft toward the $\gamma-$deformation and one can find no barrier that separates them from the prolate minima at $\beta\approx0.3$.

In order to give further insight into the relation between the mean-field models we have provided neutron and proton pairing energies of Hg isotopes in Fig. 3. The results presented here have been calculated in the same frameworks as in Fig. 1. One can see that almost all results coincide, only absolute value of proton pairing energy in the SkM$^*$-HFB model is larger.

\section{Conclusions}

We found, as expected, that various mean-field parameterizations give slightly different absolute values of the total energy. However, it is encouraging that they give quite similar dependence of the energy on the deformation. The results of this research indicate also that despite of differences between considered mean-field models the predicted pairing properties of investigated light Hg isotopes are very similar.

An important issue, not addressed in this study, is a comparison with experiment which needs a thorough consideration of various correction terms such as those resulting from the particle number or angular momentum projection. The results reported in Section 3 imply that it is necessary to include non-axial deformations in the case of considered Hg nuclei. Because of large $\gamma-$softness of the potential energy it does not seem sufficient to consider only pure prolate-oblate shapes. 

Finally, recent achievements in the self-consistent treatment of quadrupole collective excitations (in the spirit of the Generator Coordinate Method or the Adiabatic Time Dependent HFB method) open a new possibility to test the mean-field models using rich experimental data on low-lying collective states. 

\section*{Acknowledgements}
This work was supported in part by the National Nuclear Security
Administration under the Stewardship Science Academic Alliances
program through the U.S. Department of Energy Research Grant
DE-FG03-03NA00083; by the U.S. Department of Energy under Contract
Nos.\ DE-FG02-96ER40963 (University of Tennessee), DE-AC05-00OR22725
with UT-Battelle, LLC (Oak Ridge National Laboratory), and
DE-FC02-07ER41457 (UNEDF SciDAC Collaboration); by the Polish
Ministry of Science and Higher Education under Contract
N~N202~231137.


\begin{thebibliography}{0}

\bibitem{RS80}
P. Ring, P. Schuck,
{\it The Nuclear Many-Body Problem}, Springer-Verlag, Berlin, 1980.

\bibitem{Skyrme}
T. H. R. Skyrme,
{\it Nucl. Phys.} {\bf 9}, 615 (1959).

\bibitem{DG}
J. Decharg\'e and D. Gogny,
{\it Phys. Rev. C} {\bf 21}, 1568 (1980).

\bibitem{Wood92}
J. L. Wood, K. Heyde, W. Nazarewicz, M. Huyse, and P. van Duppen,
{\it Phys. Rep.} {\bf 215}, 101 (1992).

\bibitem{Andr00}
A. N. Andeyev {\it et al.},
{\it Nature} {\bf 405}, 430 (2000).

\bibitem{Julin01}
R. Julin, K. Helariutta, and M. Muikku,
{\it J. Phys. G} {\bf 27}, R109 (2001).

\bibitem{hfodd1-3}
J. Dobaczewski and J. Dudek,
{\it Comput. Phys. Commun.} {\bf 102}, 166 (1997); {\bf 102}, 183 (1997);
{\bf 131}, 164 (2000).

\bibitem{hfodd4-5}
J. Dobaczewski and P. Olbratowski,
{\it Comput. Phys. Commun.} {\bf 158}, 158 (2004); {\bf 167}, 214 (2005).

\bibitem{hfodd6}
J. Dobaczewski {\it et al.},
{\it Comput. Phys. Commun.} {\bf 180}, 2361 (2009).

\bibitem{SkM*}
J. Bartel, P.Quentin, M. Brack, C. Guet, and H.-B. H{\aa}kansson,
{\it Nucl. Phys.} {\bf A386}, 79 (1982).

\bibitem{SLy4}
E. Chabanat, P. Bonche, P. Haensel, J. Meyer, and F. Schaeffer,
{\it Nucl. Phys.} {\bf A635}, 231 (1998); {\bf A643}, 441(E) (1998).

\bibitem{pairing}
J. Dobaczewski, W. Nazarewicz, and M.V. Stoitsov,
{\it Eur. Phys. J. A} {\bf 15}, 21 (2002).

\bibitem{B}
J.-F. Berger, M. Girod, and D. Gogny,
{\it Comput. Phys. Commun.} {\bf 63}, 365 (1991).

\bibitem{C}
F. Chappert, M. Girod, and S. Hilaire,
{\it Phys. Lett. B} {\bf 668}, 420 (2008).

\bibitem{G}
S. Goriely, S. Hilaire, M. Girod, and S. P\'eru,
{\it Phys. Rev. Lett.} {\bf 102}, 242501 (2009).

\bibitem{W}
M. Warda \textit{et al.},
{\it Phys. Rev. C} {\bf 66}, 014310 (2002).

\bibitem{grahn09}
T. Grahn \textit{et al.},
{\it Phys. Rev. C} {\bf 80}, 014324 (2009).


\end{thebibliography}
\end{document}